\documentclass[twocolumn,showpacs,preprintnumbers,amsmath,amssymb]{revtex4}

\usepackage{graphicx}
\usepackage{dcolumn}
\usepackage{bm}

\begin{document}

\preprint{APS/123-QED}

\title{Negative-index media for matter-wave optics}

\author{J. Baudon}
\email{jacques.baudon@univ-paris13.fr}
\author{M. Hamamda}
\author{J. Grucker}
\author{M. Boustimi}
\altaffiliation{Present adress : Department of Physics, Umm Al-Qura University, Mekkah, Saudi Arabia}
\author{F. Perales}
\author{G. Dutier}
\author{M. Ducloy}
\affiliation{Laboratoire de Physique des Lasers, Universit\'{e} Paris 13, 93430-Villetaneuse, France}

\date{\today}

\begin{abstract}
We consider the extension of optical meta-materials to matter waves and then the down scaling of meta-optics to nanometric wavelengths. We show that the generic property of pulsed comoving magnetic fields allows us to fashion the wave-number dependence of the atomic phase shift. It can be used to produce a transient negative group velocity of an atomic wave packet, which results into a negative refraction of the matter wave. Application to slow metastable argon atoms Ar*($^3P_2$) shows that the device is able to operate either as an efficient beam splitter or an atomic meta-lens.
\end{abstract}

\maketitle
Since the pioneering work of H. Lamb \cite{Lamb} and V.G. Veselago's seminal paper \cite{Veselago} about so-called ``left-handed'' or ``meta'' media for light optics, a number of studies have been devoted to these new media and their applications (negative refraction, reversed Doppler effect, perfect lens, van der Waals atom-surface interaction, etc.) \cite{Pendry,Fotei,Sambale}, in various spectral domains \cite{Smith,Shelby,Lezec}, some of them being even extended to acoustic waves \cite{Huanyang}. Such media are essentially characterised by a negative value of the optical index, which results into opposite directions of the wave vector \textbf{k} and the Poynting vector \textbf{R}. Our goal here is to extend this concept to matter waves, and the first arising question is the following: what should be the ``de-Broglie optics'' equivalent of those meta-materials? To the energy flux in electromagnetism (\textbf{R} vector) corresponds the atomic probability flux, namely the current density of probability \textbf{J}, or equivalently the group velocity $\textbf{v}_{g}=\left|\psi\right|^{-2} \textbf{J}$, where $\psi$ is the wave-function. Therefore, here also, one has to reverse $\textbf{v}_{g}$ with respect to the wave vector \textbf{k} or the phase velocity. However, as discussed below, contrarily to what occurs in light optics where \textbf{R} remains directed outwards whereas \textbf{k} is directed towards the light source \cite{SmithKroll}, for matter waves the direction of the phase velocity (\textbf{k}) remains unchanged, whereas $\textbf{v}_{g}$ is now directed towards the source. Obviously, because of the conservation of probability, such an effect is necessarily a \textit{transient} effect. \\
In light optics, meta-media generally consist of periodic ensembles of micro- or nano- structures embedded in an ordinary material or organized into an optical band-gap crystal \cite{Lei}. Therefore their counterpart in atom optics is far from being obvious because atoms at mean and \textit{a} \textit{fortiori} low velocity (a few hundreds of m/s down to a few m/s or less) cannot penetrate dense matter. However a possible way to act on atomic waves \textit{in vacuo} is to use an interaction potential due to some external field, for instance magnetic, electric or electromagnetic fields, or the van der Waals field appearing outside a solid at the vicinity of its surface. Indeed when a semi-classical description of the external atomic motion is justified, an inhomogeneous static potential V(\textbf{r}) is equivalent to an optical index n(\textbf{r}). This comes from the fact that the optical path accumulated along a ray C (\textit{i.e.} a classical trajectory) is given by the integral $\int_{C}K(\textbf{r}) ds$, where $K(\textbf{r}) = k [1 - V(\textbf{r})/E_0]^{1/2}$ is the local wave number, k and $E_0$ being respectively the wave number and the kinetic energy of the atom in absence of potential, s is the curvilinear abscissa along the ray \cite{Landau}. This naturally leads us to set $n(\textbf{r}) = [1 - V(\textbf{r})/E_0]^{1/2}$. Nevertheless, in agreement with our previous remark about the transient character of the effect, such a type of potential cannot be a solution for our purpose because the index n is either real and positive, or purely imaginary in classically forbidden regions, but it is nowhere real negative. This looks like an \textit{impasse}. However our choice of potential was too restrictive: a much larger class of interactions is offered by position- and time-dependent potentials. Time-dependent potentials have already been widely used for devising atom optics elements in the time domain, operating either on free atom waves (modulated atomic mirrors \cite{Arndt}), or trapped atoms (pulsed interferometers \cite{Sengstock}, dynamics of Bose-Einstein condensates \cite{Andersen}). In this letter, we show that a novel class of recently introduced potentials - \textit{i.e. comoving potentials} \cite{Mathevet} - provides us with a remarkably simple solution to devise negative-index media for atomic waves.\\
Comoving potentials have been previously used in several experiments, most of them dealing with Stern-Gerlach atom interferometry \cite{Mathevet}. They have been described in detail in \cite{Mathevett}. Only the way to produce them and their main characteristics will be recalled here. Two identical planar systems of currents, periodic in space (period $\Lambda$) and symmetric with respect to $x$-axis (fig.1), produce in the vicinity of this axis a transverse field, \textit{e.g.} parallel to $y$, periodic in $x$ with the same period. Actually only the lowest spatial frequency will have a significant effect. Then the field can be assumed to be proportional to, for instance, $\cos (2 \pi x/\Lambda )$. Now if the circuit is supplied with an A.C. intensity of frequency $\nu$, then the resulting field is proportional to $\cos (2 \pi \nu (t-t_0)) \cos (2\pi  x/\Lambda ) = \frac{1}{2} [\cos (2\pi (\nu t-t_0) - x/\Lambda ) + \cos (2 \pi ( \nu( t-t_0) + x/\Lambda )]$, where $t_0$ is some reference time. Only the first term propagates in the same direction as the atoms ($z > 0$), at velocity $u =\nu \Lambda$. The period $\Lambda$  being fixed, the velocity $u$ can be varied \textit{via} $\nu$. In particular it can be tuned on the atomic velocity, giving rise to an especially important phase-shift on the atomic wave contrarily to the second counter-propagating term which can be ignored. When a continuous spectrum $H(\nu)$ is used instead of a single frequency, the resulting (pulsed) interaction potential takes the general factorized form:
\begin{equation}
V(t, x) =  s(t)\: \cos \left(2 \pi \frac{x}{\Lambda} \right)
\label{eq1}
\end{equation}
\begin{figure}
\includegraphics[width=8.5cm]{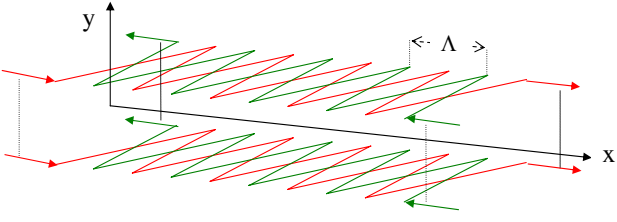}
\caption{Scheme of the device generating a transverse magnetic field ``comoving'' along the x axis. It consists of two identical planar periodic circuits (spatial period $\Lambda$). When the circuits are supplied with an A.C. current of frequency $\nu$ , they produce along $x$ a field that propagates at the velocity $u = \nu \Lambda $. When a frequency spectrum H($\nu$ ) is used, a pulsed field is generated (see text).}
\label{fig.1}
\end{figure}
The time-dependent factor is real and can be expressed as: $s(t) = g\: \mu_B M B f(t)$. Here $g$ is the atomic Land\'{e} factor, $\mu_B$ the Bohr magneton, $M$ the magnetic quantum number and $B$ the maximum of the field magnitude, $f(t)$ being a reduced dimensionless signal the maximum amplitude of which is 1. Since $s(t)$ is real, its frequency spectrum $H(\nu)$ obeys the relation:
\begin{equation}
H(\nu ) = H(-\nu )^*	
\label{eq2}
\end{equation}
where $^*$ means complex conjugate. 
The motion along the $x$ axis of a wave packet submitted to the potential $V(t, x)$ is governed by the time-dependent Schr\"{o}dinger equation:
\begin{equation}
i\hbar\:\partial_t\psi=-\frac{\hbar^2}{2m}\:\partial^2_x\psi+V(t,x)\psi	
\label{eq3}
\end{equation}
where $m$ is the atomic mass.
Taking the Fourier transform of eq. (\ref{eq3}) for variables $x, k$ ($k$ being the wave number) one gets:
\begin{equation}
i\hbar\:\partial_tC(t,k)=\frac{\hbar^2k^2}{2m}\:C+W\otimes C	
\label{eq4}
\end{equation}
where $C$ and $W$ are the spatial Fourier transforms of $\psi$ and $V$, and $\otimes$ the convolution product. Setting $C(t, k) = \exp(-i\frac{\hbar k^2}{2m} t)\Gamma(t, k)$, one readily gets:
\begin{eqnarray}
i\hbar\:\partial_t\Gamma(t, k)= s(t) \; e^{\left(\frac{i\hbar k^2}{2m}t\right)} \times \nonumber\\
\times \left(\Gamma(t, k-\kappa)e^{-i\frac{\hbar(k-\kappa)^2}{2m} t}
+\Gamma(t, k+\kappa)e^{-i\frac{\hbar(k+\kappa)^2}{2m} t}\right)
\label{eq5}
\end{eqnarray}
where $\kappa= 2\pi /\Lambda$ ($\kappa<<k$). A simplification arises because the k-derivatives of $\Gamma$ are largely dominated by those of the exponential factor $\exp(-i\frac{\hbar k^2}{2m} t)$ as soon as the wave packet has moved over a distance large compared to its own width \cite{Mathevett}. Under such conditions:
\begin{equation}
i\hbar\:\partial_t\Gamma(t,k)\approx s(t) \cos\left(2\pi\frac{\hbar k}{m \Lambda}t\right)\: \Gamma(t,k)
\label{eq6}
\end{equation}
Then:
\begin{equation}
\Gamma(t,k)\approx\Gamma(0,k)\ \exp{[i \varphi(k, t)]}
\label{eq7}
\end{equation}
where the phase shift $\varphi$ is given by :
\begin{equation}
\varphi(t,k)=-\hbar^{-1}\int^{t}_{0}dt'\ s(t') \cos\left(2\pi\frac{\hbar k}{m\Lambda}t'\right)
\label{eq8}
\end{equation}
This real phase-shift takes a limiting value at $t$ infinite, namely, assuming a perfect synchronisation of the wave packet with the field pulse (eqs. 1 and 2):
\begin{equation}
\varphi^{\infty}(k)=-\hbar^{-1}\Re[H(\frac{\hbar k}{m \Lambda})]
\label{eq9}
\end{equation}
\begin{figure}
\includegraphics[width=8.5cm]{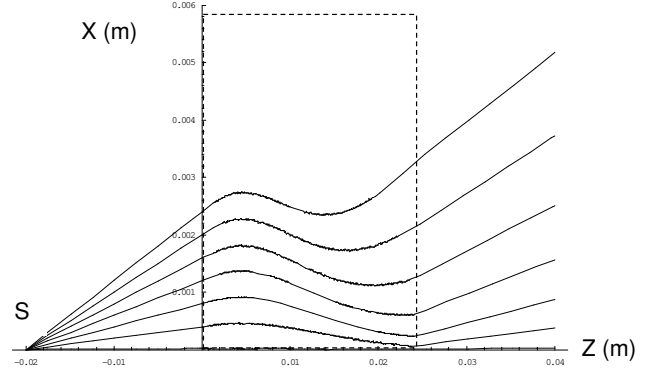}
\caption{Trajectory of the wave packet centre (coordinates $\xi(t)$, $\zeta(t)$ along $x$- and $z$-axes) in the comoving magnetic field (dashed line) propagating in the $\pm x$ direction in the region $z \geq 0$. The entrance plane is perpendicular to the $z$-axis. Rays start from a point-like source $S$ located at 2 cm from the entrance plane. These rays are in plane $x$, $z$. They initially make with the $z$ axis different incident angles ranging from 0 to 0.12 rad. All rays exhibit a \textit{negative refraction} and finally emerge, for $z > 24$ mm, parallel to their initial direction. Calculations are made for Ar*($^3P_2$, M=2) metastable atoms, the velocity of which is $v_0 = 20\ m/s$. The maximum magnitude of the magnetic field (see text) is 400 Gauss. Distances are in meters. Note the difference in scale for $x$ and $z$ axis.}
\label{fig.2}
\end{figure}
This can be seen as a \textit{genericity property} of comoving fields, in the sense that, in principle, any k-dependence of the phase-shift can be fashioned, using a convenient $H(\nu)$ spectrum. As suggested previously in \cite{Mathevett}, it could be used to balance the natural spreading of a wave packet. The factor $s(t)$, or its frequency spectrum $H(\nu )$, being given, one may derive the evolution of the wave packet in the corresponding potential. In particular, the semi-classical motion of the wave packet center is derived from the stationary phase condition: 
\begin{equation}
\partial_k [k x - \frac{\hbar k^2}{2m} t +  \varphi(k, t)] = x - \frac{\hbar k}{m} t + \partial_k  \varphi(k, t) = 0
\label{eq10}
\end{equation}
Then the abscissa of the center as a function of time is simply:
\begin{equation}
\xi(t)=\frac{\hbar k_0}{m}t-\partial_k  \varphi(k, t)|_{k_0}
\label{eq11}
\end{equation}
$\hbar k_0$ being the central momentum value. The quantity $\delta\xi(t)=-\partial_k  \varphi(k, t)|_{k_0}$ represents the spatial shift induced by the potential. It is given by:  
\begin{equation}
\delta\xi(t)=-\frac{2 \pi}{m \Lambda}\int^{t}_{0}dt'\ t' s(t') \sin\left(2\pi\frac{\hbar k}{m\Lambda}t'\right)
\label{eq12}
\end{equation}
At large values of $t$, when the potential pulse is over, this shift naturally tends to a definite limiting value, namely:
\begin{equation}
\delta\xi^{\infty}=-\frac{1}{m \Lambda} \Re[H'(\frac{\hbar k_0}{m \Lambda})]
\label{eq13}
\end{equation}
where $H'$ is the derivative of $H$.
The group velocity along $x$ is readily derived from (11) and (12):
\begin{eqnarray}
v_{gx}(t,k_0) & = & \frac{\hbar k_0}{m}-\partial_t\partial_k  \varphi(k, t)|_{k_0}\nonumber\\
 & = & \frac{\hbar k_0}{m}-\frac{2\pi}{m\Lambda}t\ s(t) \sin\left(2\pi\frac{\hbar k_0}{m\Lambda}t\right)
\label{eq14}
\end{eqnarray}
\begin{figure}
\includegraphics[width=8.5cm]{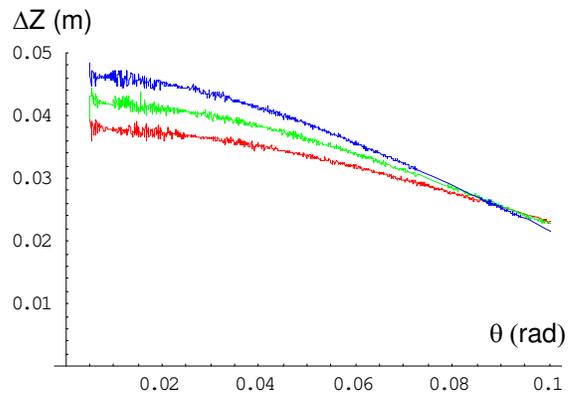}
\caption{Position of the image $S'$, where the support of the emerging ray crosses the $z$ axis. The distance $\Delta Z = SS'$ is shown (in meters) as a function of the incident angle (in rad). At $v_0 = 20$ m/s, the stigmatism is realized with an accuracy better than 90\% for incident angles lower than 0.04 rad. The chromatic dispersion is also shown (lower curve : $v_0 = 18$ m/s, mid curve : $v_0 = 20$ m/s, upper curve : $v_0 = 22$ m/s). Note the absence of chromatism at an incidence angle of 0.09 rad.}
\label{fig.3}
\end{figure}
\begin{figure}
\includegraphics[width=8.5cm]{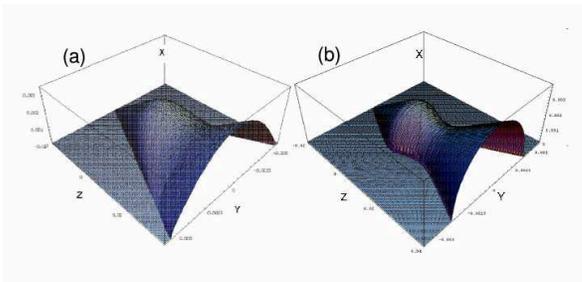}
\caption{(a) 3D representation of half a cone of rays issued from a point-like source and making an angle $\theta = 0.1$ rad with the $z$ axis. All rays exhibit a negative refraction and finally emerge parallel to their initial direction. (b) Same as (a), with a 2D comoving potential $s(t)= \left[ \cos(2\pi x/ \Lambda)+ \cos(2\pi y/ \Lambda)\right]$ (see text). The system behaves as a spherical meta-lens. Note that this surface slightly differs from that generated by the curve $\xi(z)$ rotated around the z-axis.}
\label{fig.4}
\end{figure}
The quantity  $\nu_0 =\hbar k_0/ (m\Lambda)$ can be called the ``resonance'' frequency in the sense that the velocity $u(\nu_0) = \nu_0\Lambda$  of this spectral component of the comoving field coincides with the atomic group velocity. For $t < 0$ as well as for large $t$ values, the wave packet recovers the group velocity $\hbar k_0/m$ of the free propagation. In between, it is clearly seen from (14) that, by a proper choice of $s(t)$, \textit{negative group velocities} can be (transiently) obtained whereas the phase velocity remains positive. Such behaviour is the signature of a left-handed or ``meta''-medium for atomic waves. It gives rise to new interesting effects. Let us assume that the entrance interface for atoms is the Oxy plane. Then the trajectory related to a positive magnetic quantum number ($M > 0$) is subjected to a \textit{negative refraction}, another signature of a left-handed or ``meta'' medium, whereas for $M < 0$ the trajectory undergoes an ordinary refraction, with an effective index smaller than 1. In the present case, the medium is anisotropic. It is uni-axial and optically active. Fig.2 shows rays issued from a point-like source S located at 2 cm from the interface, experiencing a negative refraction (M = +2) in the xz plane, for different incidence angles $\theta$ ranging from 0 to 0.12 rad. Numerical calculations have been carried out in the case of a nozzle beam of metastable argon atoms Ar*($^3P_2$), slowed down at a velocity $v_0 =\ 20$ m/s by means of a Zeeman slower \cite{Grucker,note19}. The parameters used in these calculations are, for the magnetic potential, $\Lambda = 5$ mm, B = 400 Gauss, $f(t) =  \epsilon^2 (t + \epsilon )^{-2} e^{-t/\tau} $ for $0\leq t\leq \tau_1$, = 0 elsewhere, with  $\epsilon =7.4$ ms,  $\tau = 0.37$ ms, $\tau_1 = 1.2$ ms. Actually the negative refraction only appears when the magnitude of B exceeds a threshold value (280 Gauss in the present case). Rays symmetric of the previous ones with respect to the $z$ axis experience opposite refractions due to the opposite component of the comoving field. The device thus behaves as a parallel plate, or a cylindrical meta-lens which gives an image S' of the source point S. In a real experimental situation, \textit{e.g.} in a cooled thermal or nozzle beam, there exists an atomic velocity distribution. The angular velocity spreading is related to the stigmatism of the system, as shown in fig.3. For an angular aperture of 0.08 rad, the axial stigmatism is limited to 3.3 mm. The dispersion of the velocity modulus gives rise to a chromatic effect. This effect is also shown in fig.3, for a wide variation ($\pm 10 \%$) of $v_{0}$. It is maximum ($\pm 4$ mm) at $\theta = 0$ and cancels at $\theta \approx 0.09$ rad. Such ``thermal'' spreadings are much larger than those experimentally accessible ($\delta\theta < 0.01$ rad, $\left|\delta v_{0}/v_{0}\right| \approx 1-2\%$). Therefore effects related to negative group velocities should be observable. Obviously ultra-cold atoms, and \textit{a fortiori}, condensates are expected to give rise to even more striking effects. The evolution of a conical map of rays with $\theta$ = 0.1 rad is shown in fig.4a. The previous treatment can be easily generalized to a 2D potential, provided it allows a separation of variables x and y, such as a potential in $\cos (2\pi x/\Lambda ) + \cos (2\pi y/\Lambda)$. Under such conditions the device behaves at small incidence angles as a spherical lens (see fig.4b). Atom wave re-focusing is actually predicted when using a series of two or three comoving pulses (Fig. 5). It is also worth noting that, with atoms initially spin-polarized in a linear superposition of Zeeman sub-levels, 1D-comoving potentials could be used in atom interferometry, as very efficient beam splitters.\\
\begin{figure}
\includegraphics[width=8.5cm]{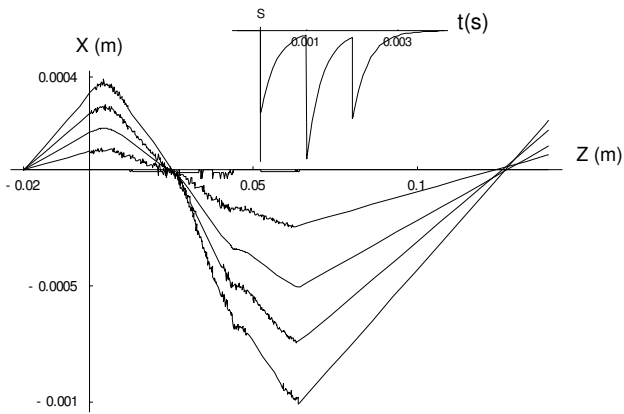}
\caption{Effect of three subsequent pulses $s(t) + 1.5 s(t - 1ms) + s(t - 2 ms)$ of comoving potential (shown in the inset), on the trajectories of the wave packet centre at incidence angles $\theta$ ranging from 0 to 16 mrad. A refocusing is seen, with a final transverse astigmatism lower than $25 \mu$m.}
\label{fig.5}
\end{figure}
	Finally, along the similarities between meta-media in light optics and atom optics, we should underline some of their basic differences. They originate in the distinct characteristics of wave propagation for Maxwell waves and ``de Broglie - Schr\"odinger'' waves: contrary to electromagnetic waves, de Broglie waves for massive particles undergo a longitudinal wave packet spreading due to the fundamental vacuum dispersion in matter optics (as imposed by the dispersion relation $\omega=\hbar k^2 /2m$). Contrarily to optical meta-materials, in \textit{scalar} de Broglie optics, atomic meta-media with negative index are characterised by a reversal of the \textit{group velocity}, the direction of the wave vector remaining unchanged. In general, retardation effects in meta-optics are important. On the other hand, in matter optics, the response time is mainly governed by the finite velocity of the massive atomic wave. In conclusion, atom optics with so-called meta-media paves the way to a new field of original operations to be performed on matter waves \cite{note20}. If the negative index for matter waves can be tuned close to $-1$, then 100\% transmission is expected at normal incidence (zero Fresnel reflection coefficient). In analogy to light wave optics \cite{Pendry}, one might be able to devise atomic lenses allowing one to re-focus atom waves without diffraction limitations. The latter characteristics should appear provided that evanescent matter waves can be properly reconstructed inside the atomic meta-medium. This key property for sub ``de-Broglie-wavelength'' focussing will be considered in a forthcoming publication. This would open novel applications in atom nano-lithography. \\
Authors are members of the \textit{Institut Francilien de Recherche sur les Atomes Froids} (IFRAF).
	

\begin{thebibliography}{99}
\bibitem{note19} Such a low velocity is also accessible using an appropriate incidence angle.
\end{thebibliography}

\end{document}